\documentclass[twocolumn,amsmath,amssymb,aps,longbibliography,floatfix]{revtex4-1}

\usepackage{cancel}
\usepackage{enumitem}
\usepackage[utf8x]{inputenc}
\usepackage{graphicx}% Include figure files
\usepackage{dcolumn}% Align table columns on decimal point
\usepackage{bm}% bold math
\usepackage{hyperref}% add hypertext capabilities
\usepackage[switch]{lineno}% Enable numbering of text and display math
\usepackage{subfigure}
\usepackage{tikz}

\usepackage{textgreek}
\usepackage{amsmath}
\usepackage{amssymb}
\usepackage{amsbsy} %bold greek symbols
\usepackage{mathrsfs}  %PER FONT \MATHSCR 
\usepackage{array,ragged2e}
\usepackage{siunitx}
\usepackage{units}%FOR THE \nicefrac COMMAND
\usepackage{braket}
\usepackage{xcolor}
\definecolor{fucsia}{RGB}{234, 89, 171}
\definecolor{celestin}{RGB}{6, 136, 235}
\usepackage{textcomp}
\usepackage{mathtools}
\usepackage{xfrac}
\usepackage{upgreek}
\usepackage{braket}
\usepackage{multirow}
\usepackage{hhline}
\usepackage{colortbl}
\usepackage[bottom]{footmisc}   
\usepackage{soul}
\usepackage{bm}
\begin{document}

\title{First-principles investigation of spin-phonon coupling in vanadium-based molecular spin qubits}

\author{$^{1}$Andrea Albino} 
\author{$^{2}$Stefano Benci}
\author{$^{1}$Lorenzo Tesi}
\author{$^{3}$Matteo Atzori}
\author{$^{2}$Renato Torre}
\author{$^{4}$Stefano Sanvito}
\author{$^{1}$Roberta Sessoli}
\email{roberta.sessoli@unifi.it}
\author{$^{4}$ Alessandro Lunghi}
\email{lunghia@tcd.ie}

\affiliation{$^{1}$ Dipartimento di Chimica "Ugo Schiff" and INSTM RU, Universit\'a degli Studi di Firenze, I50019 Sesto Fiorentino, Italy}
\affiliation{$^{2}$ Dipartimento di Fisica ed Astronomia and European Laboratory for Nonlinear Spectroscopy, Universit\'a degli Studi di Firenze, I50019 Sesto Fiorentino, Italy}
\affiliation{$^{3}$ Dipartimento di Chimica "Ugo Schiff" and INSTM RU, Universit\'a degli Studi di Firenze, I50019 Sesto Fiorentino, Italy. Present address: Laboratoire National des Champs Magnetiques Intenses, UPR 3228 - CNRS, F38042 Grenoble, France.}
\affiliation{$^{4}$ School of Physics, CRANN and AMBER, Trinity College, Dublin 2, Ireland}

\begin{abstract}
{\bf Paramagnetic molecules can show long spin-coherence times, 
which make them good candidates as quantum bits. Reducing the efficiency of the spin-phonon 
interaction is the primary challenge towards achieving long coherence times over a wide temperature range 
in soft molecular lattices. The lack of a microscopic understanding about the role of vibrations in spin relaxation strongly undermines 
the possibility to chemically design better performing molecular qubits. Here we report a first-principles characterization of the main 
mechanism contributing to the spin-phonon coupling for a class of vanadium(IV) molecular qubits. Post Hartree Fock and 
Density Functional Theory are used to determine the effect of both reticular and intra-molecular vibrations on the modulation 
of the Zeeman energy for four molecules showing different coordination geometries and ligands. This comparative study 
provides the first insight into the role played by coordination geometry and ligand field strength in determining the spin-lattice 
relaxation time of molecular qubits, opening the avenue to a rational design of new compounds.}
\end{abstract}

\maketitle

\section{Introduction}

Quantum information science deals with the representation, storage and processing of information by means of a 
quantum mechanical system. The basic element is the quantum bit or \textit{qubit}, namely the quantum analogous 
of the classical bit ($0,1$).
Quantum computation exploits the quantum properties of the qubit, such as superposition and 
entanglement, \cite{Acin2018} providing an ideal platform for improving algorithms' efficiency. In 
particular, one can design a range of quantum algorithms, which scale with the complexity of the 
problem in a much more favourable way than their classical counterparts.
Several systems are currently investigated for the practical realization of quantum devices. For example: superconducting circuits, \cite{Devoret2013} trapped ions \cite{Cirac1995}
 and polarized photons.\cite{Ringbauer2017} 
Among the various physical systems with potential for developing quantum technologies the \textit{spin}, with its intrinsic two-levels qubit structure, occupies
a special place. 

Both nuclear spins\cite{Morton2008} and electron spins,\cite{Leuenberger2001} as well as electron-nuclear hybrid 
systems,\cite{Hussain2018,Carretta2018} can be exploited for this purpose. Electron spins interact more strongly with the 
environment, compared to the nuclear ones, and thus they are easier to read out. In contrast, electrons have 
shorter spin lifetimes and must be carefully protected from the environment, while keeping the possibility to interact 
with each others. In practice three physical systems implement spin qubits: nitrogen-vacancy centres 
in diamond (NVC),\cite{Hanson2008} atomic impurities in semiconductors, such as P implanted in Si,\cite{Bowyer2015}
and paramagnetic molecules.\cite{Ding2016, Harneit2017, Warner2013} In contrast to solid-state spin qubits based 
on dopant atoms, such as the NVCs, molecules show a significant advantage, namely the chemical systems hosting 
the spin can be tailored to tune the quantum properties and the coupling to other qubits. This can create quantum 
platforms,\cite{Ferrando-Soria2016,Aguila2014,Carretta2018} providing a bottom-up route to large-scale quantum 
register fabrication.

A spin lifetime at least $10^4$ times longer than the time needed for an individual quantum operation\cite{Shor1994,Preskill2012} 
represents the minimal requirement for the development of a qubit. Accordingly, the figures of merit to consider in the 
design of electronic spin molecular qubits are fundamentally two:\cite{Tesi2016} $\mathit{i}$) the longitudinal (or 
\textit{spin-lattice}\cite{Abragam1970}) relaxation time, $T_1$, which corresponds to the lifetime of a classical 
bit; $\mathit{ii}$) the coherence (or \textit{spin-spin} relaxation) time, $T_2$, which is the time
characteristic for a spin to loose memory of a coherent quantum superposition state. In the last few years, remarkable 
results have been achieved against both $T_1$ and $T_2$ by investigating mononuclear transition-metal complexes.\cite{Zadrozny2014,Warner2013,Graham2014,Zadrozny2017,Andres2018,Bader2014,Bader2016,Tesi2016,Tesi2016a,Atzori2016,Atzori2016a} These results place molecules containing light metals \textit{back in the quantum race}.~\cite{Sessoli2015}
Vanadium(IV)-based complexes represent a promising class of compounds to be used as fundamental 
components in quantum technologies. These spin 1/2 systems show long spin-spin relaxation time $T_2$, up to $1$ 
millisecond at liquid helium temperature, when complexed with nuclear-spin free ligands and diluted in nuclear-spin free solvents.\cite{Zadrozny2015}
This property makes V$^\mathrm{IV}$-based compounds very attractive for further development. 

A common trend found in molecular spins\cite{Atzori2017c,Atzori2016} is the rapid decrease of $T_2$ on raising the 
temperature, a feature that limits their potential use at room temperature. The interaction with lattice vibrations, typically 
connected to the $T_1$-type relaxation, also contributes to the $T_2$-type one and becomes the predominant relaxation 
mechanism when increasing temperature.\cite{Bader2014} In solids, thermal motion is usually described by phonons, 
which are energy quanta of lattice vibrations. Spin-lattice relaxation is caused by the absorption/emission of phonons by 
the spin system. This process is possible due to the presence of spin-orbit coupling,\cite{Gill1975} an interaction that couples 
atomic and spin degrees of freedom and enables the energy exchange among the two systems.

Although the experimental investigations in this field are numerous,\cite{Atzori,Atzori2016a,Atzori2017c,Tesi2016,Tesi2016a}
the theoretical description is still at an early stage.\cite{Escalera-Moreno2017,Cardona-Serra2018,Lunghi2017,Lunghi2017a,Ferrando-Soria2016} 
In particular, the possibility to include molecules in a solid ab initio computational framework has been made possible 
only in the last few years, enabled by an extensive work of integration of density functional theory (DFT), post Hatree 
Fock (postHF) methods and spin dynamics for the calculation of the dynamical magnetic properties of multi-spin systems.

\begin{figure*}[t]
	\begin{tikzpicture}	
	[auto]
	\draw[help lines,color=white
	] (0,0) grid (18,4);
	\node at (2,2.5) (1) 
	{\includegraphics[scale=1]{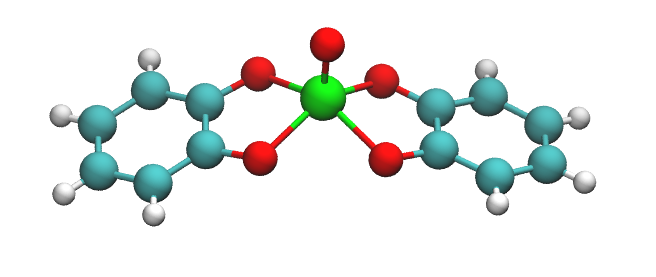}};
	\node at (6.5,2.5) (2) {\includegraphics[scale=1]{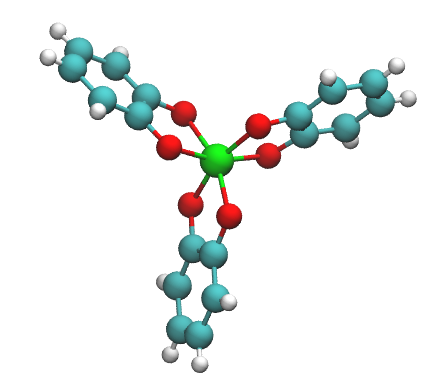}};
	\node at (10.9,2) (3) {\includegraphics[scale=1]{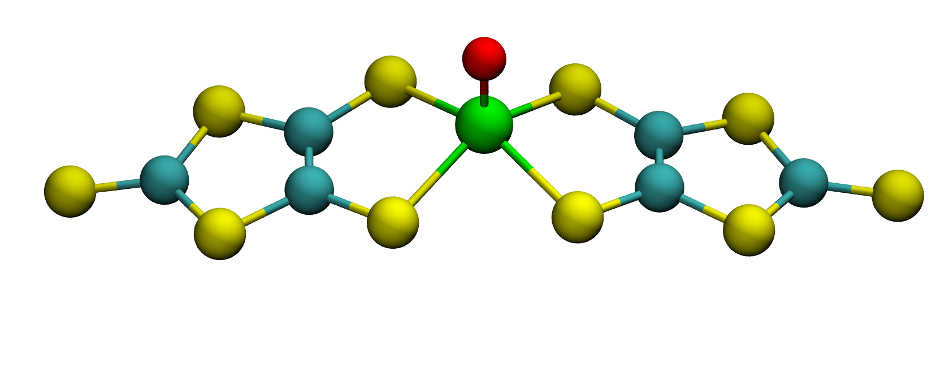}};
	\node at (16.1,2) (4)
	{\includegraphics[scale=1]{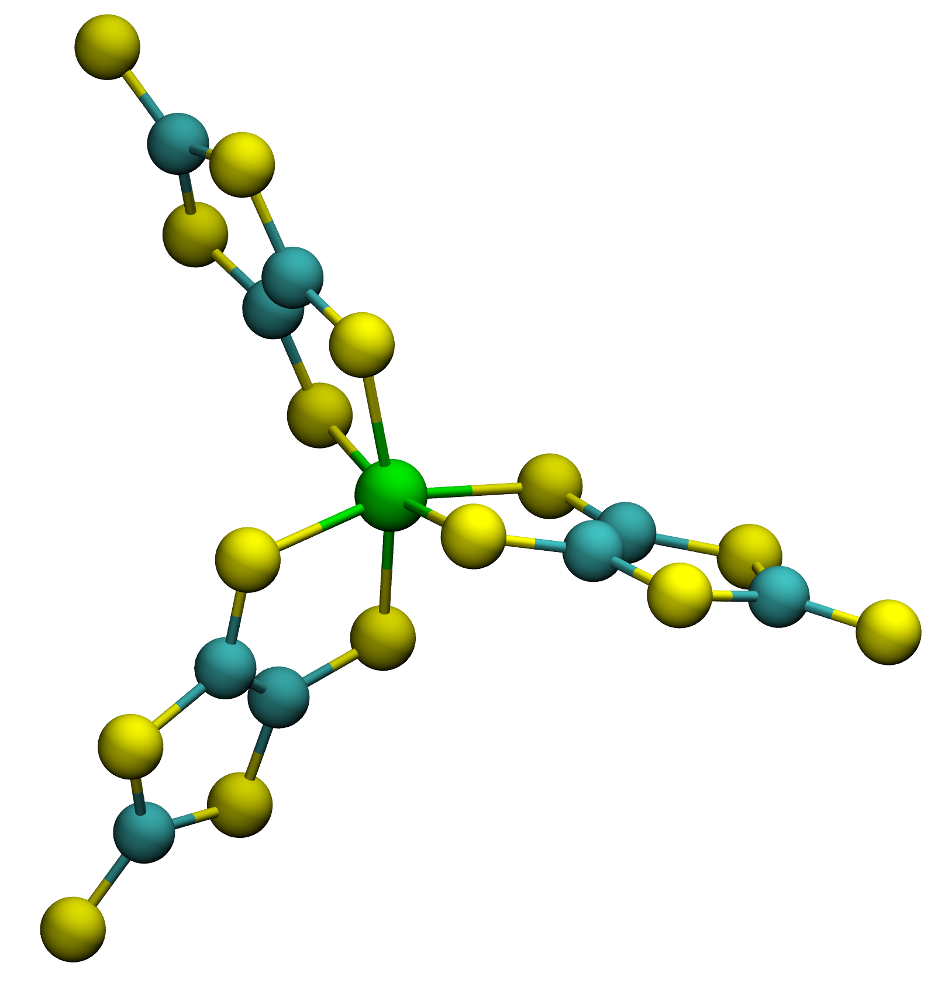}}; 
	\node at (2,0) (1b) {\textbf{1}. \sffamily
		[VO(cat)$_2$]\textsuperscript{2-}};
	\node at (6.5,0) (2b) {\textbf{2}. \sffamily  [V(cat)$_3$]\textsuperscript{2-}};
	\node at (10.9,0) (3b) {\textbf{3}. \sffamily [VO(dmit)$_2$]\textsuperscript{2-}};
	\node at (16.1,0) (4b) {\textbf{4}. \sffamily
		[V(dmit)$_3$]\textsuperscript{2-}};
	
	\node at (16,2.7) (4b) {\sffamily \footnotesize S$_2$};
	\node at (15.1,2.3) (4b) {\sffamily \footnotesize S$_1$};
	\node at (16.6,2.3) (4b) {\sffamily \footnotesize S$_3$};
	\node at (16.2,1.4) (4b) {\sffamily \footnotesize S$_4$};
	\node at (14.8,1.9) (4b) {\sffamily \footnotesize S$_6$};
	\node at (16,1) (4b) {\sffamily \footnotesize S$_5$};
	
	\node at (5.75,2.75) (4b) {\sffamily \footnotesize O$_1$};
	\node at (6.3,3.7) (4b) {\sffamily \footnotesize O$_2$};
	\node at (7,3.55) (4b) {\sffamily \footnotesize O$_3$};
	\node at (7,2.55) (4b) {\sffamily \footnotesize O$_4$};
	\node at (7,2) (4b) {\sffamily \footnotesize O$_5$};
	\node at (5.75,2.25) (4b) {\sffamily \footnotesize O$_6$};
	
	\node at (10.4,2.95) (4b) {\sffamily \footnotesize S$_2$};
	\node at (10.4,1.45) (4b) {\sffamily \footnotesize S$_1$};
	\node at (11.6,2.85) (4b) {\sffamily \footnotesize S$_3$};
	\node at (11.6,1.45) (4b) {\sffamily \footnotesize S$_4$};
	
	\node at (1.5,2) (4b) {\sffamily \footnotesize O$_1$};
	\node at (1.5,3.25) (4b) {\sffamily \footnotesize O$_2$};
	\node at (2.6,3.2) (4b) {\sffamily \footnotesize O$_3$};
	\node at (2.6,2) (4b) {\sffamily \footnotesize O$_4$};
	
	\end{tikzpicture} 
	\caption{Molecular structures of the dianionic complexes \textbf{1}-\textbf{4} with the first coordination 
	sphere atom labelling
		scheme.}
	\label{molequle}
\end{figure*}
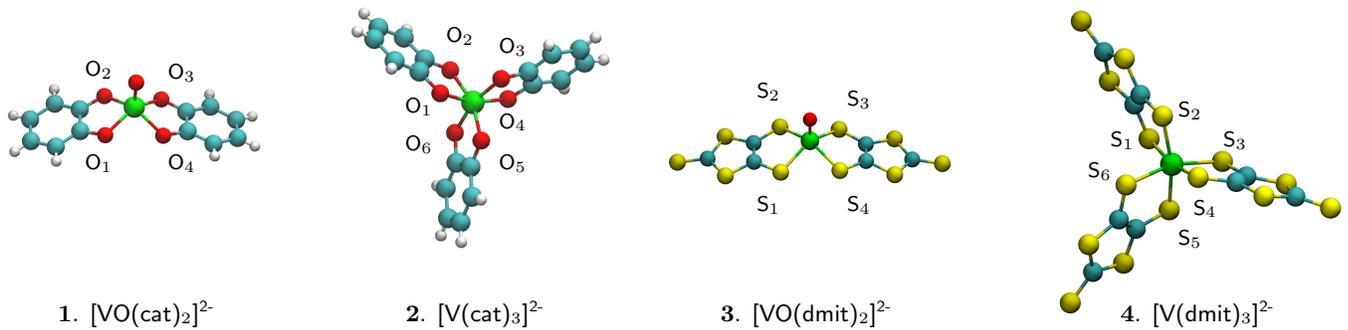 

The present work introduces a comparative theoretical investigation of the spin-lattice relaxation in four V$^\mathrm{IV}$
molecular complexes (\autoref{molequle}). In particular, we have focused on penta-coordinated vanadyl 
(VO\textsuperscript{2+}) and hexa-coordinated V$^\mathrm{IV}$ molecules, where the coordination is obtained 
by cathecolate\cite{Atzori2017c} and dithiolene\cite{Atzori2016a} ligands, namely 
[PPh$_4$]$_2$[VO(cat)$_2$] (\textbf{1}), 
[PPh$_4$]$_2$[V(cat)$_3$] (\textbf{2}), 
[PPh$_4$]$_2$[VO(dmit)$_2$] (\textbf{3}), 
[PPh$_4$]$_2$[V(dmit)$_3$] (\textbf{4}) (cat = catecholate, dmit = 1,3-dithiole-2-thione-4,5-dithiolate, PPh$_4$ = tetraphenylphosphonium). 
These complexes have already experimentally shown to possess long coherence times and remarkable 
differences in the temperature dependence of the spin-lattice relaxation time. Our approach consists in 
modelling their magnetic properties by first-principles, when perturbing the molecular structures along the normal modes 
of vibration, following a strategy adopted in previous works.\cite{Escalera-Moreno2017,Lunghi2017} In order to correctly account for the vibrational properties of solid state systems, where intermolecular 
interactions become relevant,
DFT calculations are performed in the crystal phase, in contrast to previous studies where the gas phase was taken into 
consideration.\cite{Escalera-Moreno2017} The quality of our computed vibrational properties is ascertained by IR vibrational 
spectroscopy in the THz range. The information extracted from DFT and post Hartee-Fock methods together provide a fingerprint 
description of the interaction between vibrations and magnetism, and directly correlate to the structure of each compound. 
The correlation found between the spin-phonon interaction amplitudes and experimental spin-lattice relaxation times 
for the four compounds is then discussed. 

\section{Methods}

\subsection{Vibrational properties calculations}
The simulations of the crystals' vibrational properties are performed with a Gaussian 
and plane waves (GPW) formalism as implemented in the Quickstep module\cite{VandeVondele2005a} of the 
CP2K\cite{Hutter2014,cp2k} package. The GGA\cite{Perdew2001} functional in the \textit{Perdew-Burke-Ernzerhof} 
approximation\cite{Perdew1996} (PBE) is chosen for all the calculations. Van-der-Waals interactions are taken into 
account with the non local rVV10 correction scheme.\cite{Sabatini2013} The calculation of the Hessian matrix is 
performed by finite differences, the Hessian matrix is symmetrized averaging it with its transpose and the acoustic sum rule is applied to impose the translational invariance.\cite{Califano1981} Each atomic coordinate is displaced by $\pm0.01 \SIUnitSymbolAngstrom$ 
around the equilibrium configuration, and such displacements are used to compute energies and forces.
When displacing the atoms we have taken into account the crystal's symmetries and only inequivalent 
displacements have been considered (see Figure S1 in Supplementary Information - SI). The exploitation of 
symmetry allows us to reduce the computational overheads by a factor of four for all the systems considered.
The diagonalization of the Hessian matrix provides the phonons frequencies, $\omega_{\alpha}$ (Table S1 in SI), 
and the normal modes of vibrations, $q_{\alpha}$. The latter are stored in the columns of the Hessian's eigenvectors 
matrix $\mathbf{L}$.

\subsection{Spin-phonon coupling coefficients calculations}

The calculation of the spin and the spin-phonon Hamiltonian parameters has been carried out with 
the ORCA\cite{Neese2017,Neese2018} package. The level of theory used is Complete Active Space 
Self Consistent Field plus second order perturbation theory (CASSCF+NEVPT2), with a def2-TZVP basis 
set for V, O and S, whereas def2-SVP is used for C and H. The active space includes one electron and 
the five $d$-orbitals of the molecule. The molecular geometry used for these simulations is obtained by the 
periodic DFT calculation for the optimized crystal cell.

The calculation of the spin-phonon coupling coefficients is performed following a tensor differentiation procedure 
as described in a previous report on Single Molecule Magnets (SMMs).\cite{Lunghi2017} This procedure is here 
applied to the Land\'e $\mathbf{g}$ tensor that describes the coupling between the spins and an external magnetic 
field. Each element of the Land\'e tensor of the equilibrium structure (indicated by the subscript 0) is differentiated 
with respect to the $3M$ \textit{atomic Cartesian positions} $\left( \partial \mathbf{g}/\partial \mathbf{X}\right)_0$, 
where $M$ is the number of atoms in the molecule, instead of the $3N$ unit-cell vibrational coordinates, $q_{\alpha}$, 
\cite{Lunghi2017a} (see Figure S2 in SI). 
This approach requires a number of $\mathbf{g}$-derivatives equal to $3M=78$, instead of $3N=1392$ for 
[VO(cat)$_2$]\textsuperscript{2-}, and  $3M=111$ instead of $3N=1524$ for [V(cat)$_3$]\textsuperscript{2-} 
(see \autoref{cristalli}). The $\mathbf{g}$ tensor is calculated for six displaced geometries 
($\pm0.0050 \SIUnitSymbolAngstrom$, $\pm0.0075 \SIUnitSymbolAngstrom$, 
$\pm0.0150 \SIUnitSymbolAngstrom$) around the equilibrium configuration for the $3M$ molecular coordinates.
The $\mathbf{g}$-versus-displacement curves are then fitted to a second order polynomial expression (see Figs.~S3
and S4 in the SI for a demonstration of the fit quality).

The Cartesian derivatives of the $\mathbf{g}$ tensor are used to compute two important parameters: the average 
molecular spin-phonon coupling, $|\partial\mathbf{g}|$, and the phonon-projected spin-phonon coupling coefficients, 
$\left(\partial \mathbf{g}/\partial q_{\alpha}\right)_{0}$. The formers are defined as,
\begin{equation}
|\partial\mathbf{g}|=\sum_{\mathit l\mathit v}^{M,3}\sum_{\mathit j \mathit r}^{3} \left|\Big(\frac{\partial g_{\mathit j \mathit r}}{\partial X_{\mathit l \mathit v}}\Big)_{0}\right|\:,
\end{equation}
where the index $\mathit l$ runs over the number of atoms in the molecule and $\mathit v$ runs over the cartesian coordinates.
The phonons projected spin-phonon coupling coefficients are instead defined as
\begin{equation}   
\Big(\frac{\partial \mathbf{g}}{\partial q_{\alpha}}\Big)_{0}=\sum_{i}^{3M}\sqrt{\frac{\hbar}{\omega_{\alpha}m_{\mathit i}}}L_{\mathit i\alpha}\Big(\frac{\partial \mathbf{g}}{\partial X_{\mathit i}}	\Big)_{0}\:,
\end{equation}
where the index $\alpha$ runs over the normal modes, the index $i$ over the 3$M$ molecular degrees of freedom and 
$L_{i\alpha}$ is the Hessian's eigenvectors matrix. 

\subsection{Experimental}

Compounds \textbf{1}-\textbf{4} were prepared as described in the literature.\cite{Atzori2016a,Atzori2017c}
 THz spectra were measured by time-domain transmission spectroscopy using a table-top experimental set-up equipped with optical laser pulses (T-light 780 nm fiber laser, MenloSystems) and low-temperature GaAs photoconductive antennas. Low temperatures measurements are allowed by means of a closed-cycle Helium cryostat in the temperature range from 10 to 300 K. The developed acquisition procedure enables to achieve a signal-to-noise ratio higher than what is commonly achieved in standard far-infrared investigations. The accurate analysis of the data enables to disentangle the signals from spurious contributions coming from multiple reflections. The detailed description of the experimental set-up and of the material parameters extraction procedure (i.e. absorption coefficient, refractive index) is reported elsewhere.\cite{Tasseva2017, Taschin2018}
  The spectra were measured in pellets of 13.2 mm diameter and thickness of about 0.7 mm. These were made by pressing under a manual hydraulic press ($\sim$ 0.8 GPa) a mixture of microcrystals and polyethylene powder (Merck).

\section{Results}
\subsection{Spin-phonon dynamics theory}

\begin{table*}
	\centering
	\small
	\caption{Collection of structural data obtained from single crystal X-ray diffraction\cite{Zadrozny2015,Atzori2016a,Cooper1982} and from theoretical calculations.}
	\label{cristalli}
	\begin{tabular}{@{}lrlrlrlrl@{}}	
		\hline
		\multirow{1}{*}{} 
		& \multicolumn{2}{c}{\textbf{1}}
		& \multicolumn{2}{c}{ \textbf{2}}
		& \multicolumn{2}{c}{ \textbf{3}}
		& \multicolumn{2}{c}{ \textbf{4}}\\
		\multirow{1}{*}{\qquad\textsc{Molecule}} 
		& \multicolumn{2}{c}{$\left[\text{VO(cat)$_2$}\right]$\textsuperscript{2-}}
		& \multicolumn{2}{c}{$\left[\text{V(cat)$_3$}\right]$\textsuperscript{2-}}
		& \multicolumn{2}{c}{$\left[\text{VO(dmit)$_2$}\right]$\textsuperscript{2-}}
		& \multicolumn{2}{c}{$\left[\text{V(dmit)$_3$}\right]$\textsuperscript{2-}}\\
		\hline
		Counter-ion 
		&\multicolumn{2}{c}{ 2$\times$[PPh$_4$]\textsuperscript{+}}
		&\multicolumn{2}{c}{ 2$\times$[PPh$_4$]\textsuperscript{+}}
		&\multicolumn{2}{c}{ 2$\times$[PPh$_4$]\textsuperscript{+}}
		&\multicolumn{2}{c}{ 2$\times$[PPh$_4$]\textsuperscript{+}}\\
		
		Molecule Atoms ($M$)   
		&\multicolumn{2}{c}{26}
		&\multicolumn{2}{c}{37}
		&\multicolumn{2}{c}{18}
		&\multicolumn{2}{c}{25}
		\\

		Crystal Cell Atoms ($N$)
		&\multicolumn{2}{c}{464}
		&\multicolumn{2}{c}{508}
		&\multicolumn{2}{c}{432}
		&\multicolumn{2}{c}{460}\\
		
		Crystal System 
		&\multicolumn{2}{c}{monocline} 
		&\multicolumn{2}{c}{monocline}
		&\multicolumn{2}{c}{monocline}
		&\multicolumn{2}{c}{monocline} \\
		
		Spatial Group 
		&\multicolumn{2}{c}{P21/c}
		&\multicolumn{2}{c}{C2/c}
		&\multicolumn{2}{c}{C2/c}
		&\multicolumn{2}{c}{P21/c}
		\\
		
		Site symmetry (Z)
		&\multicolumn{2}{c}{4}
		&\multicolumn{2}{c}{4}
		&\multicolumn{2}{c}{4}
		&\multicolumn{2}{c}{4}\\

		&exp.&sim.&exp.&sim.&exp.&sim.&exp.&sim.\\
		Cell Volume, \si{\angstrom}$^3$
		& 4734.95 & 4477.07
		& 5015.83 & 4758.34
		&5346.43  & 4941.81
		&5744.23  & 5428.57\\
		a, \si{\angstrom} 
		&13.25& 12.95
		&15.31& 14.99
		&20.47& 19.62
		&24.57& 24.10 \\
		
		b, \si{\angstrom}
		&12.25& 11.95
		&13.23& 13.16
		&12.73& 12.29
		&13.81& 13.69\\
		
		c, \si{\angstrom} 
		&29.19& 28.96
		&25.32& 24.55
		&20.60& 20.53
		&18.13& 17.56\\
		
		$\beta$, deg. 
		&92.80&  92.58
		&102.02& 100.87
		&95.29&  93.33
		&111.01& 110.39
		\\
		V--L, \si{\angstrom} (av.)
		&1.973& 1.974
		&1.946& 1.959
		&2.387& 2.382
		&2.386& 2.373
		\\ 
		V=O, \si{\angstrom} (av.)
		&1.614& 1.640
		&--& --
		&1.594& 1.621
		&--& --
		\\
		\hline
	\end{tabular}
\end{table*} 

If we limit ourselves to the study of spin dynamics in the presence of an 
external magnetic field, $\bm{B}$, the spin Hamiltonian of an $|\bm{S}|$=1/2 system will only contain the 
Zeeman term,
\begin{equation} \label{hamspin}
\mathscr{H}_{\text{s}} = \mu_\mathrm{B} \bm{B}\cdot\mathbf{g}\cdot\bm{S}\:,
\end{equation}
where $\mu_\mathrm{B}$ is the Bohr magneton and $\mathbf{g}$ is the Land\`e tensor. The hyperfine and spin-spin 
dipolar interactions have been neglected in \autoref{hamspin} because in high fields their matrix elements are negligible. 

When dealing with relaxation properties, it is necessary to introduce in the description the effects of 
the environment on the dynamics of the system described. For this we need an open quantum systems
formalism.\cite{Petruccione2002} Here the spin system interacts with an environment made of the crystal's 
phonons. Their Hamiltonian, describing the normal modes of vibration, is obtained as the second-order Taylor 
expansion of the nuclei potential energy surface and reads
\begin{equation}
\mathscr{H}_{\text{ph}} = \sum_{\alpha}\hbar\omega_{\alpha}(n_{\alpha} + \frac{1}{2}),
\end{equation} 
where $n_{\alpha}=a^{\dagger}a$ is the phonon density operator, $a^{\dagger}$ and $a$ the creation and annihilation operators.
In first approximation, assuming a weak coupling between the phonons bath and the 
spin degrees of freedom, the spin-phonon coupling Hamiltonian can be
taken as linear in the ionic displacement
\begin{equation}\label{pertt}
\mathscr{H}_{\text{s-ph}} = \sum_{\alpha}\left( \frac{\partial\mathscr{H}_{\text{s}}}{\partial q_{\alpha}}\right)_0 q_{\alpha}\:.  
\end{equation}

The spin dynamics can then be described by the Redfield equations,\cite{Redfield1957} where the reduced spin 
density matrix, $\rho^\mathrm{S}(t)$, evolves in time because of the interaction with phonons
\begin{align}\label{redfield} \nonumber
\frac{d\rho_{\mathit a\mathit a}^\mathrm{S}(t)}{dt} &= \frac{2}{\hslash^2}\sum_{\alpha}\sum_{\mathit b}\mathscr{M}^{\alpha}_{\mathit{a}\mathit{b}}\rho_{\mathit{bb}}^\mathrm{S}(t),\\ 
\mathscr{M}^{\alpha}_{\mathit{a}\mathit{b}}=
%\left\lbrace
-\sum_{\mathit j}V_{\mathit{aj}}^{\alpha}V_{\mathit{jb}}^{\alpha}&G(\omega_{\mathit{jb}},\omega_{\alpha})+|V_{\mathit{ab}}^{\alpha}|^2G(\omega_{\mathit{ba}},\omega_{\alpha}),
%\right\rbrace.
\end{align}
where  $G(\omega_{\mathit{ij}},\omega_{\alpha})$ is the Fourier transform of the phonon correlation function 
and it is defined as
\begin{align}
G(\omega_{\mathit{ij}},\omega_{\alpha}) &= 
\delta(\omega_{\mathit{ij}}-\omega_{\alpha})\bar{n}_{\alpha}+\delta(\omega_{\mathit{ij}}+\omega_{\alpha})(\bar{n}_{\alpha}+1)\:.
\end{align}
Here $\bar{n}_{\alpha}=1/(e^{\frac{\hslash\omega_{\alpha}}{kT}}-1)$ is the phonon occupation number, 
according to Bose-Einstein statistics, and 
$V_{\mathit{ab}}^{\alpha}=\braket{\mathit a|\frac{\partial \mathscr{H}_{\text{s}}}{\partial q_{\alpha}}|\mathit b}$ 
is the matrix element of the spin part of the spin-phonon coupling Hamiltonian, $\mathscr{H}_{\text{s-ph}}$, 
evaluated between two eigenfunctions, $\ket{\mathit a}$ and $\ket{\mathit b}$, of $\mathscr{H}_{\text{s}}$. 
The approximations involved in this approach have been discussed elsewhere.\cite{Lunghi2017} 
Understanding the interactions contributing to $V_{\mathit{ab}}^{\alpha}$ is the main focus of this work. 

In order to understand the origin of the spin-phonon coupling is necessary to discuss the nature of 
the spin Hamiltonian that enters in \autoref{pertt}. The anisotropy of the spin Hamiltonian through the 
$\mathbf{g}$ tensor represents the fingerprint of the spin-lattice interaction, which is mediated by the 
presence of the spin-orbit coupling.\cite{Gatteschi,Neese1998} Let us show explicitly the contribution 
of the $\mathbf{g}$ tensor to the spin-phonon coupling Hamiltonian
\begin{align}\label{ya}
 \mathscr{H}_{\text{s-ph}} =\sum_{\alpha}\sum_{\mathit j\mathit r}\mu_{B}\hat{S}_{\mathit j}B_{\mathit r}\left( \frac{\partial g_{\mathit j\mathit r}}{\partial q_{\alpha}}\right)_0 q_{\alpha}.
\end{align}
Estimating $\mathscr{H}_{\text{s-ph}}$ thus requires the calculation of the derivatives of 
$\mathbf{g}$ with respect to the structural perturbations and the calculation of the periodic crystal's 
normal modes.

\subsection{Structural and vibrational properties}

\begin{figure*}[t]
	\centering
	\begin{tikzpicture}	
	[auto]
	\draw[help lines,color=white
	] (0,0) grid (18,3.5);
	\node at (11.3,1.7) (1) {\includegraphics[scale=1]{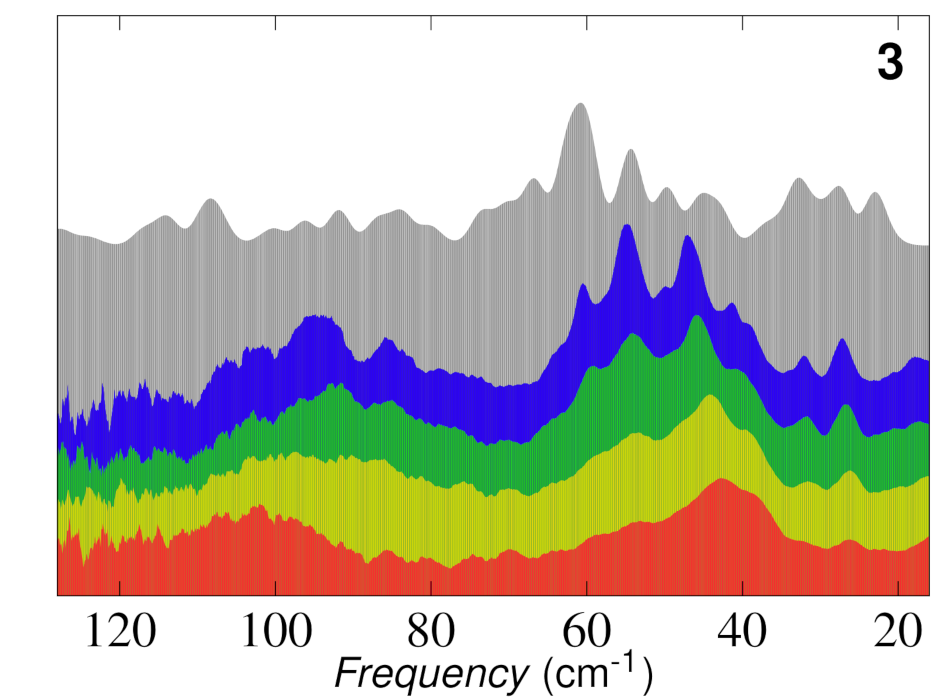}};
	\node at (16,1.7) (2) {\includegraphics[scale=1]{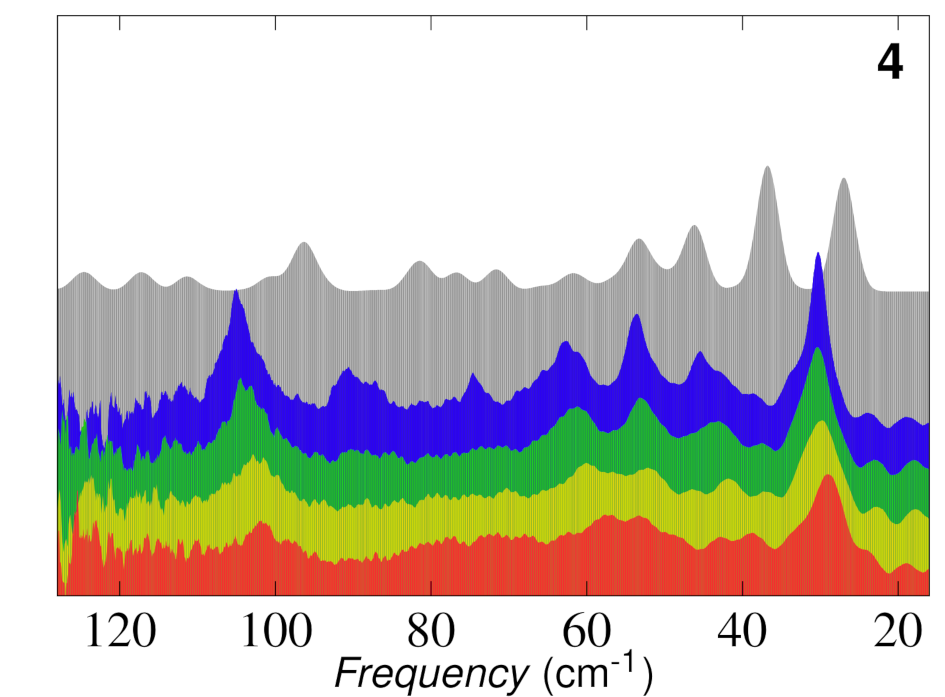}};
	\node at (1.93,1.7) (3) {\includegraphics[scale=1]{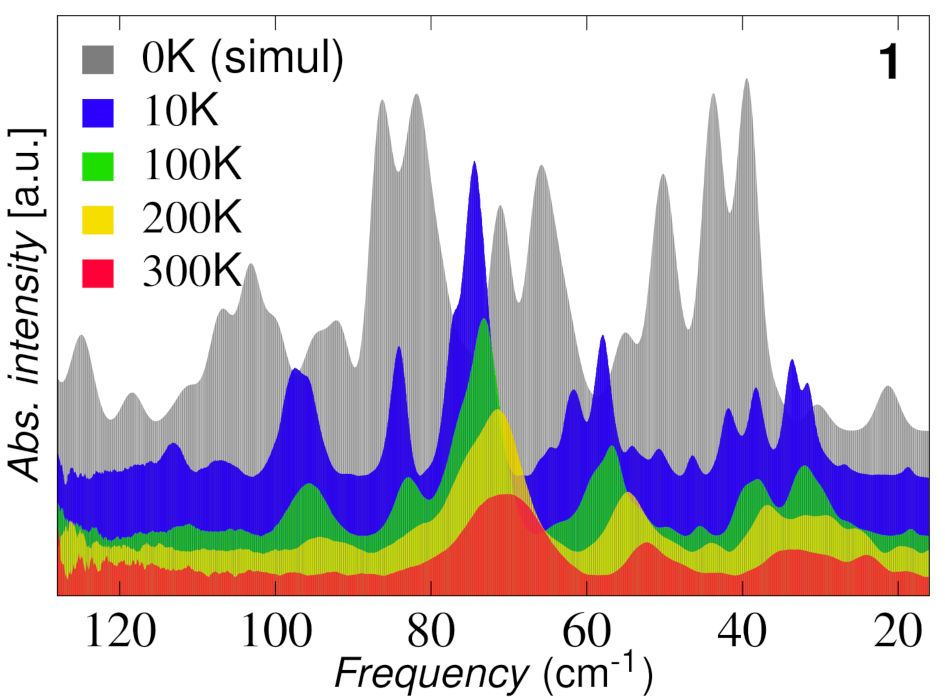}};
	\node at (6.61,1.7) (4) {\includegraphics[scale=1]{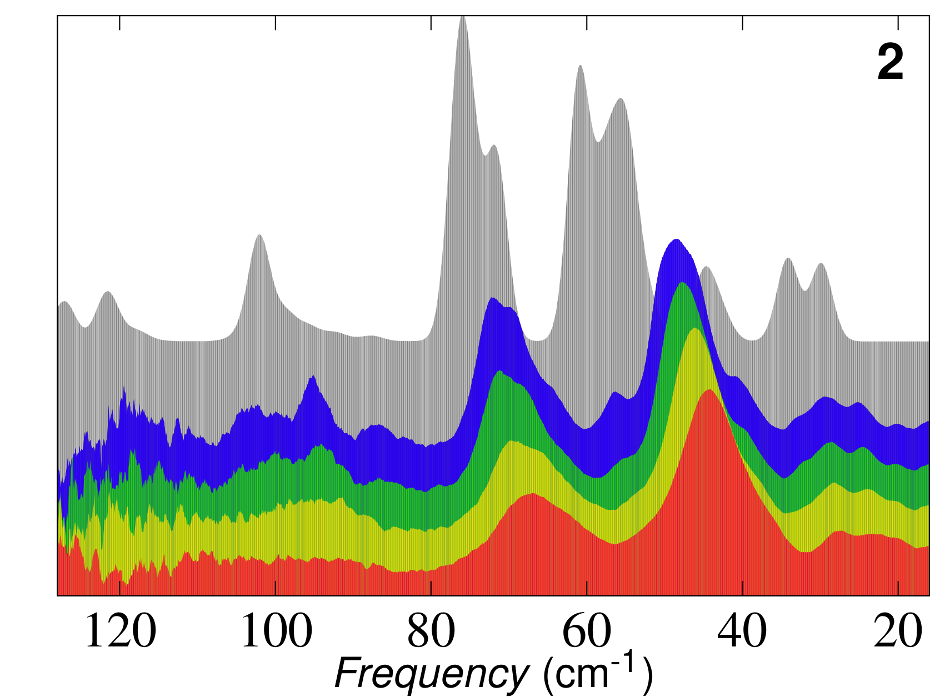}};
	\end{tikzpicture} 
	\caption{Experimental and simulated THz spectra for compounds \textbf{1}-\textbf{4}. The former are measured 
	at several temperatures between 10 and 300 K for powder-like samples embedded in polyethylene.}
	\label{tiaccazeta}
\end{figure*} 

The procedure outlined in the Computational Methods section has been employed for the \textbf{1}-\textbf{4} 
crystal structures, whose details are summarized in Table \ref{cristalli} and Figure S5 in the SI. 
The coordination geometries around the metal centres are square pyramidal, for the penta-coordinate compounds 
($\mathbf{1}$,$\mathbf{3}$), and trigonally-distorted octahedral for the hexa-coordinate ones ($\mathbf{2}$,$\mathbf{4}$).
It is worth noting that the metal--ligand (V--L) distances in sulphur-containing ligands are longer than those of the oxygen ones (\autoref{cristalli}), consistent with the difference in atomic radius 
between O and S.

The starting point for understanding the effects of the first coordination shell on the spin-relaxation properties 
is the analysis of the crystal vibrations. One crystallographic cell ($\Gamma$-point approximation) is
optimized and the Hessian matrix is calculated numerically. This gives $3N-3$ optical modes. Limiting our study to the $\Gamma$-point, the $3$ acoustic 
modes have all $\omega_\alpha=0$.\cite{Califano1981}

Experimental far-infrared (IR) THz spectroscopy as a function of temperature  with a spectral window between 15 to 120~cm\textsuperscript{-1}  
is employed here to support the quality of our simulations for complexes \textbf{1}-\textbf{4} (see Fig.~\ref{tiaccazeta}). 
The lowest temperature spectrum ($10$ K) shows an overall good agreement with 
our simulated spectra. A temperature increase causes a red-shift of some vibrational modes because of the presence 
of anharmonic interactions together with the softening of the crystal lattice. Simulations corresponding to 
$0$~K, indeed, show blue-shift with respect to the experimental lowest-temperature spectra.
The first calculated vibrations occur at $13.3$ cm\textsuperscript{-1} for \textbf{4} and $18.5$ cm\textsuperscript{-1} 
for \textbf{3}, while at $27.6$ cm\textsuperscript{-1} for \textbf{2} and $20.7$ cm\textsuperscript{-1} for \textbf{1}.  Dithiolene 
compounds show vibrations at lower frequencies with respect to the cathecolate ones, probably 
because of the larger radius of first coordination sphere and the higher atomic weight of sulfur. Accordingly, 
longer bond lengths are generally associated to softer bonds and, therefore, to lower vibrational frequencies. 

\begin{figure}[!h]
	\begin{tikzpicture}	
	[auto,
	line/.style={draw, thick, black,shorten >=0pt},
	line2/.style={draw, thick, -latex,shorten >=-6pt},
	line3/.style={draw, semithick, black, decoration={brace, raise=0.0pt, %mirror,
			amplitude=1mm}, decorate}]

	\draw[help lines, color=white
	] (0,0) grid (9,7);
	
	\node at (2.35,5.25) (3) {\includegraphics[scale=1]{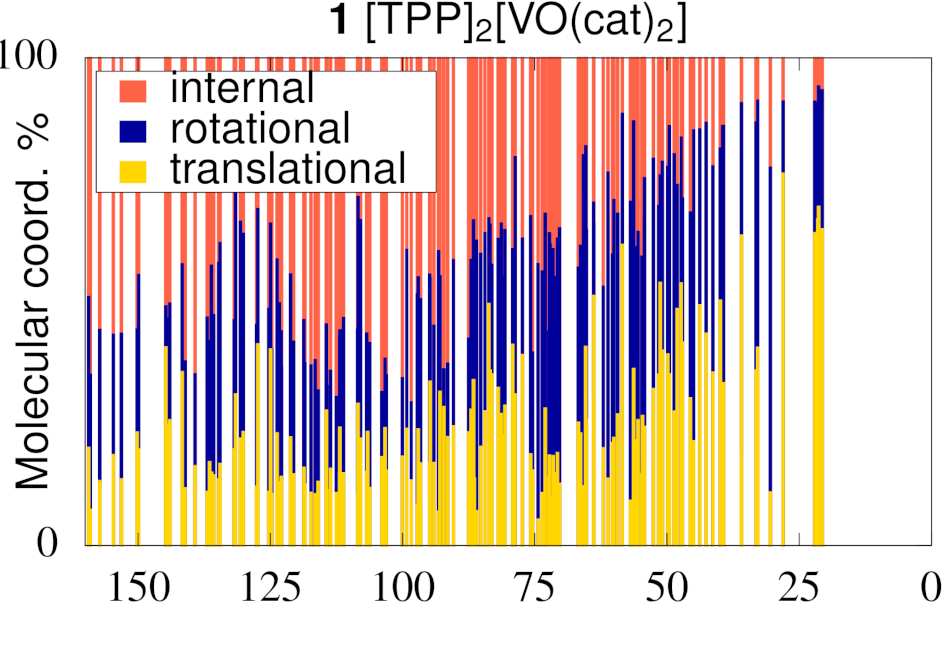}};
	\node at (6.55,5.25) (4) {\includegraphics[scale=1]{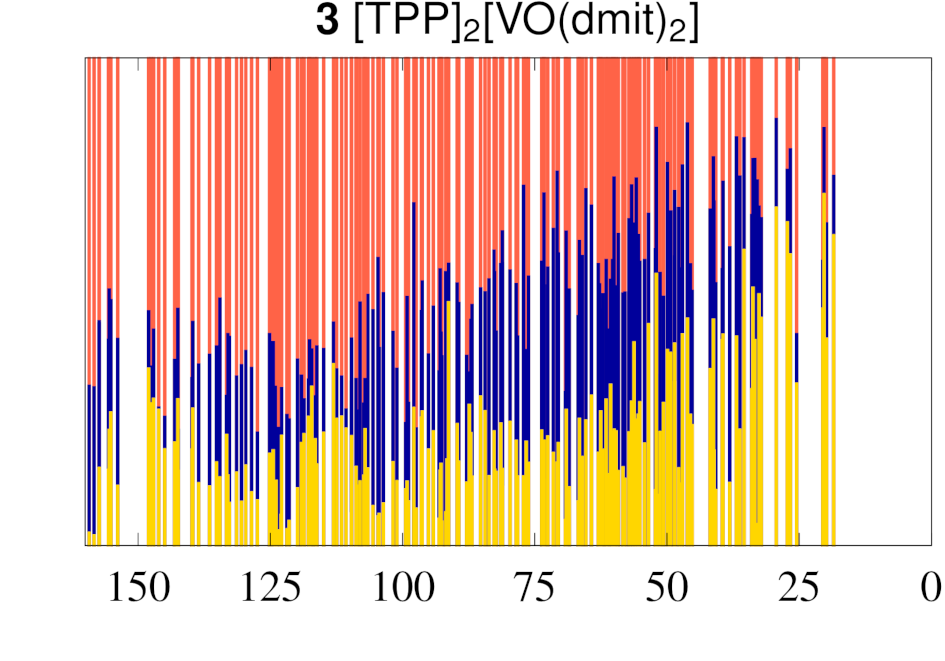}};
	\node at (2.35,2.2) (3) {\includegraphics[scale=1]{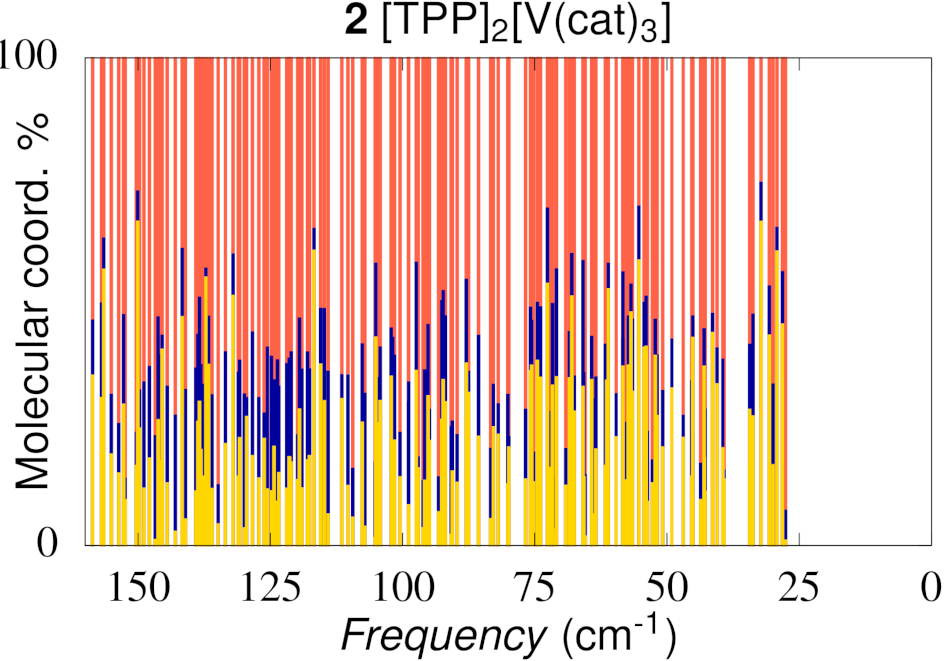}};
	\node at (6.55,2.2) (4) {\includegraphics[scale=1]{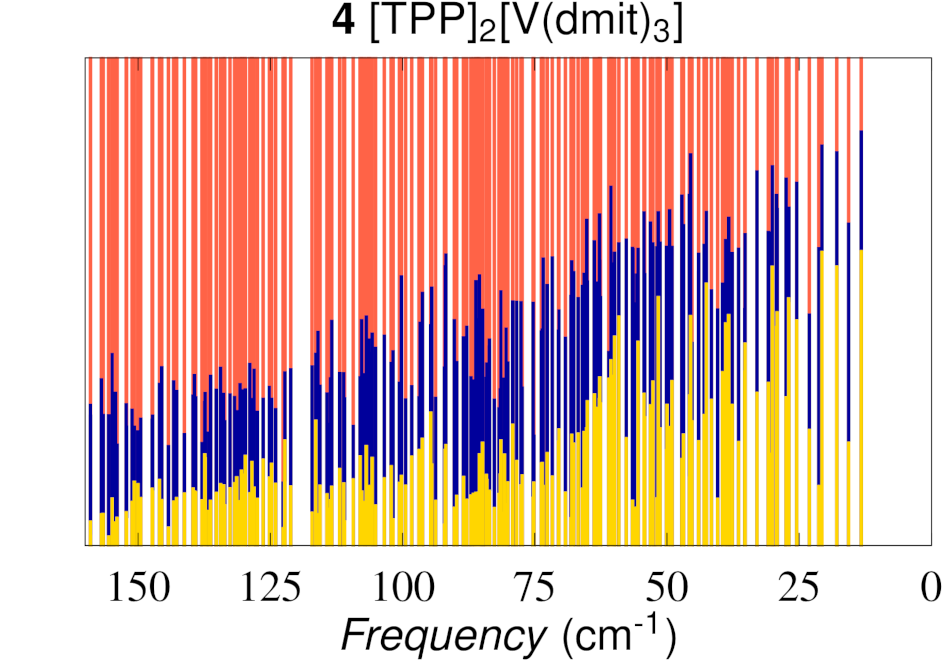}};

	\end{tikzpicture} 
	\caption{The total molecular displacement associated to normal modes with frequency in the range 
	0--150 cm\textsuperscript{-1} is decomposed in intra-molecular motion (red), motion associated 
	to molecular rotations (blue) and motion associated to molecular translations (yellow).}
	\label{dekomp}
\end{figure}

A deconvolution of the vibrational modes in molecular translations, molecular rotations and intra-molecular 
motions, has been performed following the method outlined in previous works,\cite{Neto2006,Califano1981} 
and the results are shown in \autoref{dekomp}. The low-energy modes are dominated, for all compounds, 
by rigid translations and rotations of the molecule in the crystal, but internal contributions are also present 
and become the dominant ones on increasing the modes energy. The calculated decomposition shows, for 
hexa-coordinate compounds, an higher average internal contribution. Among the hexa-coordinate, the cathecolate 
compound shows a reduced rotational contribution (see blue lines in Fig.~\ref{dekomp}). 

\subsection{Spin-phonon coupling analysis }

\autoref{epierre} shows the calculated and experimental $\mathbf{g}$-values. The deviation of $\mathbf{g}$ from 
the free electron value of 2.0023 is caused by the presence of spin-orbit coupling. In
general, a good agreement between experimental and theory is observed and the trend 
in magnitude among the different molecules is correctly reproduced. \autoref{epierre} also reports the averaged 
molecular spin-phonon coupling coefficients, as defined in the Computational Methods section. These parameters represent the
sum of every atomic contribution to the spin-phonon coupling and allow us to define dynamical 
magneto-structural correlations. It should be noted that the differences in the average spin-phonon coupling across the series can 
be due to both the chemical nature of the ligand and the coordinating geometry around the metal centre. For complexes with the 
same ligand, the $\mathbf{g}$ tensor elements are less perturbed in the pyramidal coordination with respect to the octahedral one. 
When comparing complexes with the same coordination geometry but different ligands, it appears that a stronger 
effect on the spin-phonon coupling is observed, with the cathecolate being more prone to $\mathbf{g}$ tensor perturbations 
than the dithiolene ones.

\begin{table} [!h]
	\small
	\centering
	\caption{Best fit parameters extracted from simulation of the experimental CW-EPR X-band (ca. $9.7$ GHz) spectra and 
	CASSCF simulation results from optimized equilibrium geometry of $\mathbf{1}$-$\mathbf{4}$.\cite{Atzori2016a,Atzori2017c}}
	\label{epierre}
	\begin{tabular}{@{}lcccc@{}}	
		\hline
		& \textbf{1}
		& \textbf{2}
		& \textbf{3}
		& \textbf{4}\\
		&
		$\left[\text{VO(cat)$_2$}\right]$\textsuperscript{2-}&
		$\left[\text{V(cat)$_3$}\right]$\textsuperscript{2-}&
		$\left[\text{VO(dmit)$_2$}\right]$\textsuperscript{2-}&	
		$\left[\text{V(dmit)$_3$}\right]$\textsuperscript{2-}\\
		\hline
		&\multicolumn{4}{c}{Experimental Parameters (room T)}\\
		$g_x$& 1.980(1)
		& 1.945(1)
		& 1.986(1)
		& 1.961(1) \\
		$g_y$& 1.988(1)
		& 1.945(1)
		& 1.988(1)
		& 1.971(1)\\
		$g_z$& 1.956(1)
		& 1.989(2)
		& 1.970(1)
		& 1.985(1)\\
		\hline
		&\multicolumn{4}{c}{Simulated Parameters ($0$K)}\\		 
		$g_x$& 1.978(1)
		& 1.906(1)
		& 1.987(1)
		& 1.943(1)\\	
		$g_y$&1.984(1)
		& 1.914(1)
		& 1.988(1)
		& 1.956(1)\\			
		$g_z$& 1.935(1)
		& 1.997(1)
		& 1.953(1)
		& 1.997(1)\\
		\hline
		&\multicolumn{4}{c}{Average Spin-Phonon Coupling}\\
	        $|\partial \mathbf{g} |$ & 1.5456 & 6.1460 & 0.3838 & 2.5774 \\
		\hline
	\end{tabular}		
\end{table} 

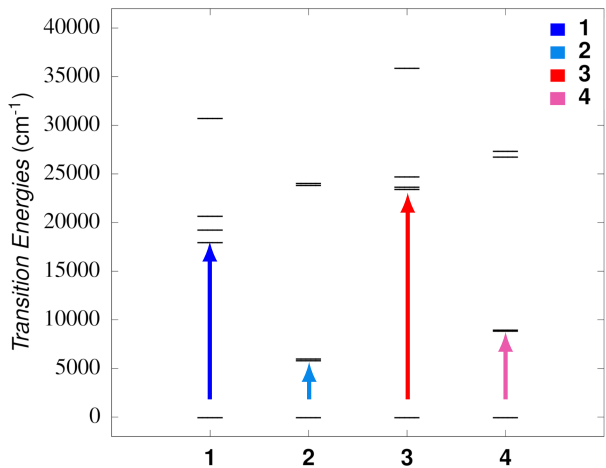
\begin{figure}[h]
	\begin{tikzpicture}	
	[auto,
	line/.style={draw, ultra thick, -latex, black,shorten >=0pt},
	line2/.style={draw, thick, -latex,shorten >=-6pt},
	line3/.style={draw, semithick, black, decoration={brace, raise=0.0pt, %mirror,
			amplitude=1mm}, decorate}]
	\draw[help lines,color=white
	] (0,0) grid (9,7);
	\node at (4.5,3.5) (1) {	\includegraphics[scale=1]{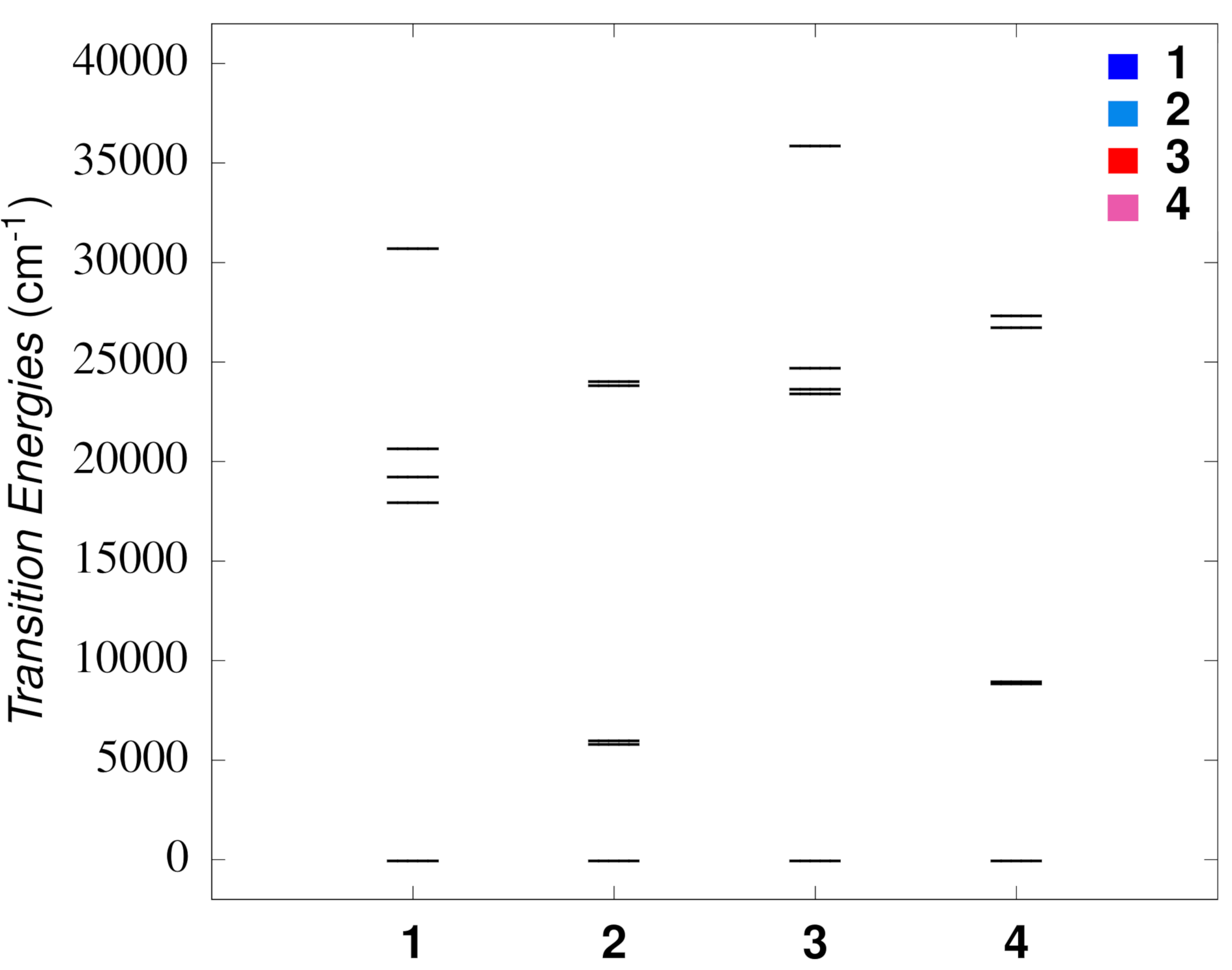}};
	
	\node at (3.17,1.3) (uno){};
	\node at (3.17,3.4) (due){};
	\node at (4.49,1.3) (tre){};
	\node at (4.49,1.8) (qua){};
	\node at (5.8,1.3) (cin){};
	\node at (5.8,4.05) (sei){};
	\node at (7.1,1.3) (set){};
	\node at (7.1,2.2) (ott){};
	
	\begin{scope}[every path/.style=line, overlay ]
	\path[color=blue] (uno.center) --  (due.center);
	\path[color=red] (cin.center) -- (sei.center);
	\path[color=celestin] (tre.center) -- (qua.center);
	\path[color=fucsia] (set.center) -- (ott.center);
	\end{scope}
	\end{tikzpicture} 
	\caption{CASSCF + NEVPT2 calculated energy ladder of the 3$d$ valence shell of the four compounds analysed.}
	\label{struttura}
\end{figure}

In order to explain these features it is important to correlate the $\mathbf{g}$-tensor anisotropy and 
how this is modified by atomic displacements, i.e. $\mathbf{g}$ and $|\partial \mathbf{g} |$. These two quantities show the same trend across the series of molecules 
investigated as a consequence of a common microscopic origin, namely the magnitude of orbital angular 
momentum in the ground state.\cite{Lever} This quantity, accessible from our \textit{ab initio} calculations, 
can be conveniently estimated from the magnitude of the 3d orbitals energy 
splitting,\cite{Lever} showed in \autoref{struttura} for the compounds $\mathbf{1}$-$\mathbf{4}$. The first excited-state splitting, 
calculated by CASSCF+NEVPT2, shows a good correlation with both the $\mathbf{g}$ factors (\autoref{epierre}) and their 
derivatives, where larger $\mathbf{g}$-shifts and $\mathbf{g}$ derivatives correspond to smaller $\Delta E$. Interestingly, 
oxygen ligands yield a smaller splitting of the electronic states with respect to the sulphur ones. 

\begin{figure}[!h]
	\begin{tikzpicture}	
	[auto,
	line/.style={draw, thick, black,shorten >=0pt},
	line2/.style={draw, thick, -latex,shorten >=-6pt},
	line3/.style={draw, semithick, black, decoration={brace, raise=0.0pt, %mirror,
			amplitude=1mm}, decorate}]

	\draw[help lines,color=white
	] (0,0) grid (9,5.8);

	\node at (2.25,3) (3) {\includegraphics[scale=1]{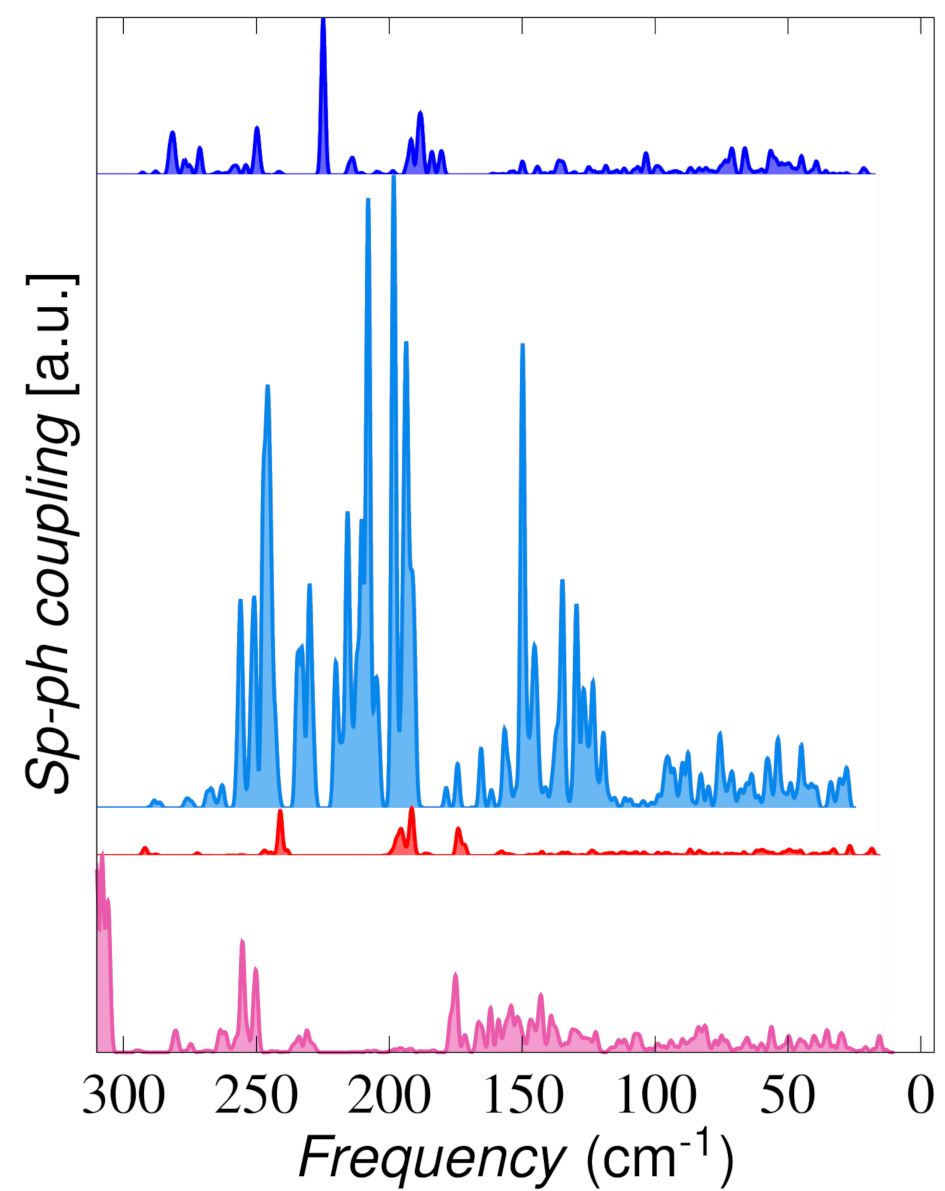}};
	\node at (6.75,3) (4) {\includegraphics[scale=1]{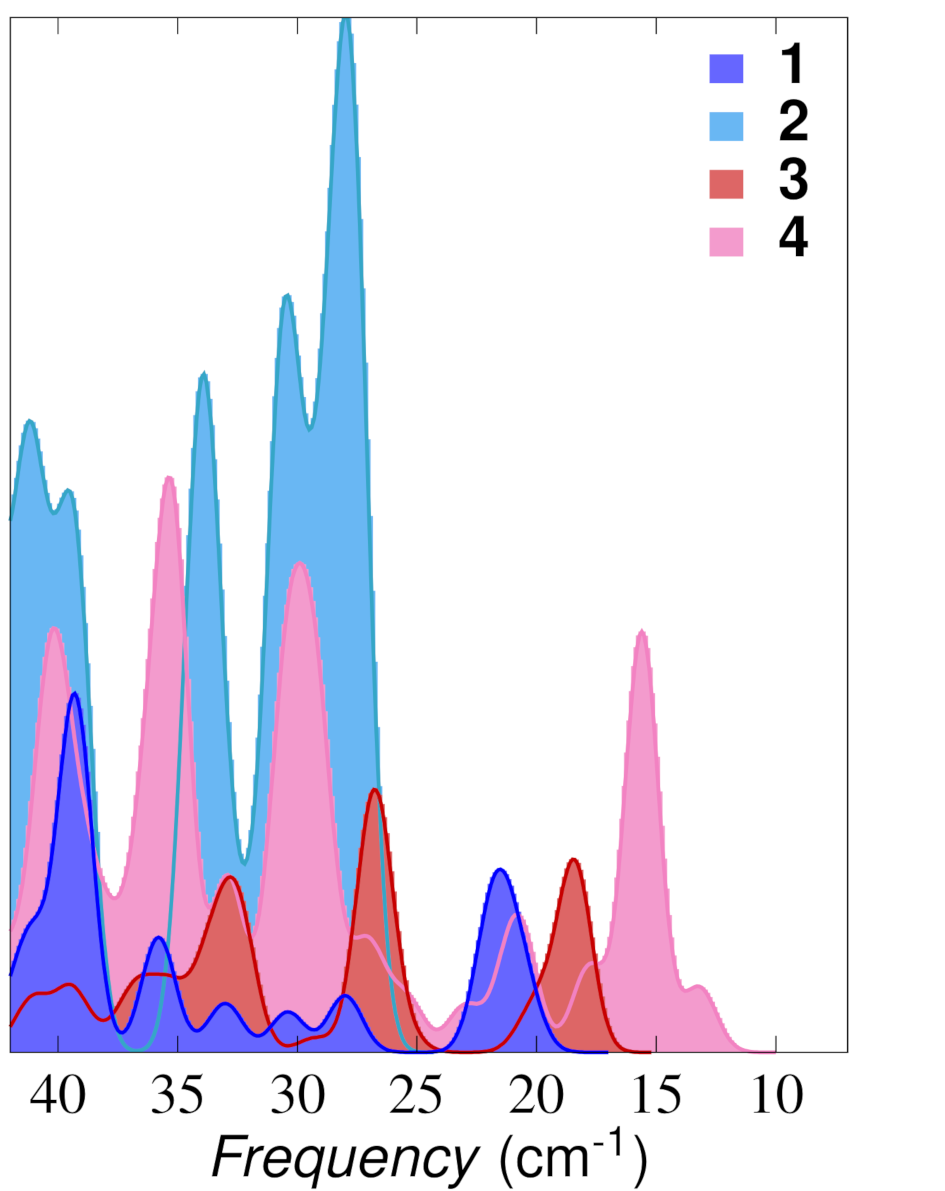}};

	\node at (3.93,5.12) (uno){};
	\node at (4.4,5.12) (due){};
	\node at (4.4,0.85) (tre){};
	\node at (3.93,0.85) (qua){};
	\node at (4.54,0.9) (cin){};

	\node at (0.3,5.45) (){\textbf{a}};
	\node at (4.5,5.45) (){\textbf{b}};

%	\begin{scope}[every path/.style=line, overlay ]
%	\path (uno.center) --  (due.center);
%	\path (due.center) -- (tre.center);
%	\path (tre.center) -- (qua.center);
%	\path (qua.center) -- (uno.center);
%	\end{scope}
%	\begin{scope}[every path/.style=line2, overlay ]
%	\path[-] (tre.west) edge [bend right] (cin);
%	\end{scope}
	\end{tikzpicture} 
	\caption{Left panel: spin-phonon coupling coefficients in the 0-300 cm\textsuperscript{-1} range for the four analysed 
	compounds. A Gaussian line shape was applied to each harmonic normal mode, considering a width parameter equal to 2 cm$^{-1}$. 
	Right panel: superposition of the first modes coupling coefficients with same Gaussian line broadening.}
	\label{spinfonone}
\end{figure}
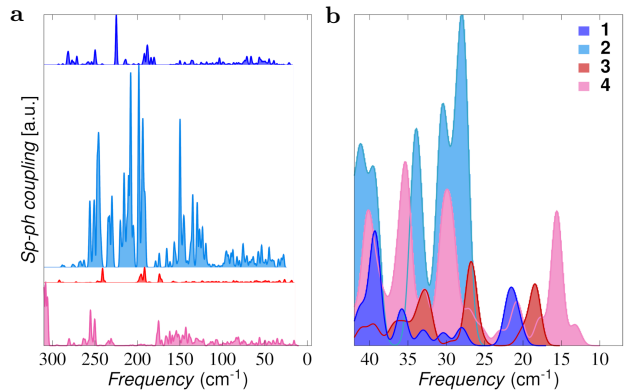

Although useful to understand general trends, the average spin-phonon coupling coefficients $|\partial \mathbf{g}|$ 
do not provide any information concerning the temperature at which specific atoms will start vibrating. 
This information can be obtained from the study of the spin-phonon coupling coefficients projected on the normal 
modes, as displayed in Figure \ref{spinfonone}a and \ref{spinfonone}b as a function of the phonons' vibrational frequency. 
The overall behaviour of the spin-phonon coupling follows the one observed for $|\partial \mathbf{g}|$, that is the strength 
of the coupling and ranks the molecules in the following order $\mathbf{3}<\mathbf{1}<\mathbf{4}<\mathbf{2}$. Thus, 
the spin-phonon coupling in vanadyl compounds is weaker than that in hexa-coordinated molecules. Furthermore, 
cathecolate ligands offer a stronger coupling than dithiolenes.
%%%%%%
The presence of different donor atoms has different effects on the low and high energy ranges of the vibrational 
spectrum. Indeed, dithiolene ligands, which exhibit a more diffuse coordination sphere due to the softer nature of S, show 
$\mathit{i})$ the presence of several normal vibrations at lower frequencies, $\mathit{ii})$ a weaker perturbation of the spin states by 
vibrational modes, as shown by the lower spin-phonon coupling amplitudes (see Table S1 in SI). The second effect 
is visible in the higher energy vibrations, suggesting a ligand dependence of $\tau$ vs. temperature.

\section{Discussion}

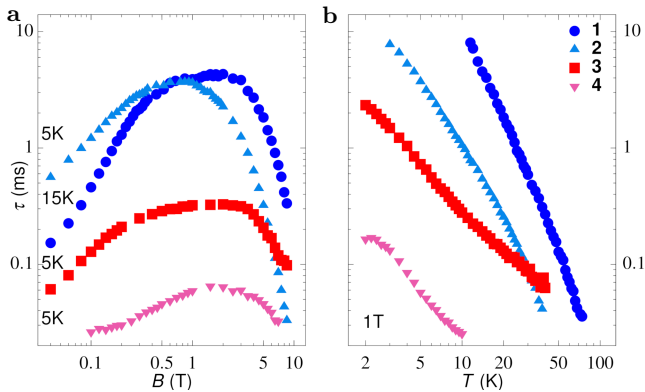
\begin{figure}[!h]
	\begin{tikzpicture}	
	[auto,
	line/.style={draw, thick, black,shorten >=0pt},
	line2/.style={draw, thick, -latex,shorten >=-6pt},
	line3/.style={draw, semithick, black, decoration={brace, raise=0.0pt, 
			amplitude=1mm}, decorate}]
	
	\draw[help lines,color=white
	] (0,0) grid (9,5.8);
	
	\node at (2.21,3) (1) {\includegraphics[scale=1]{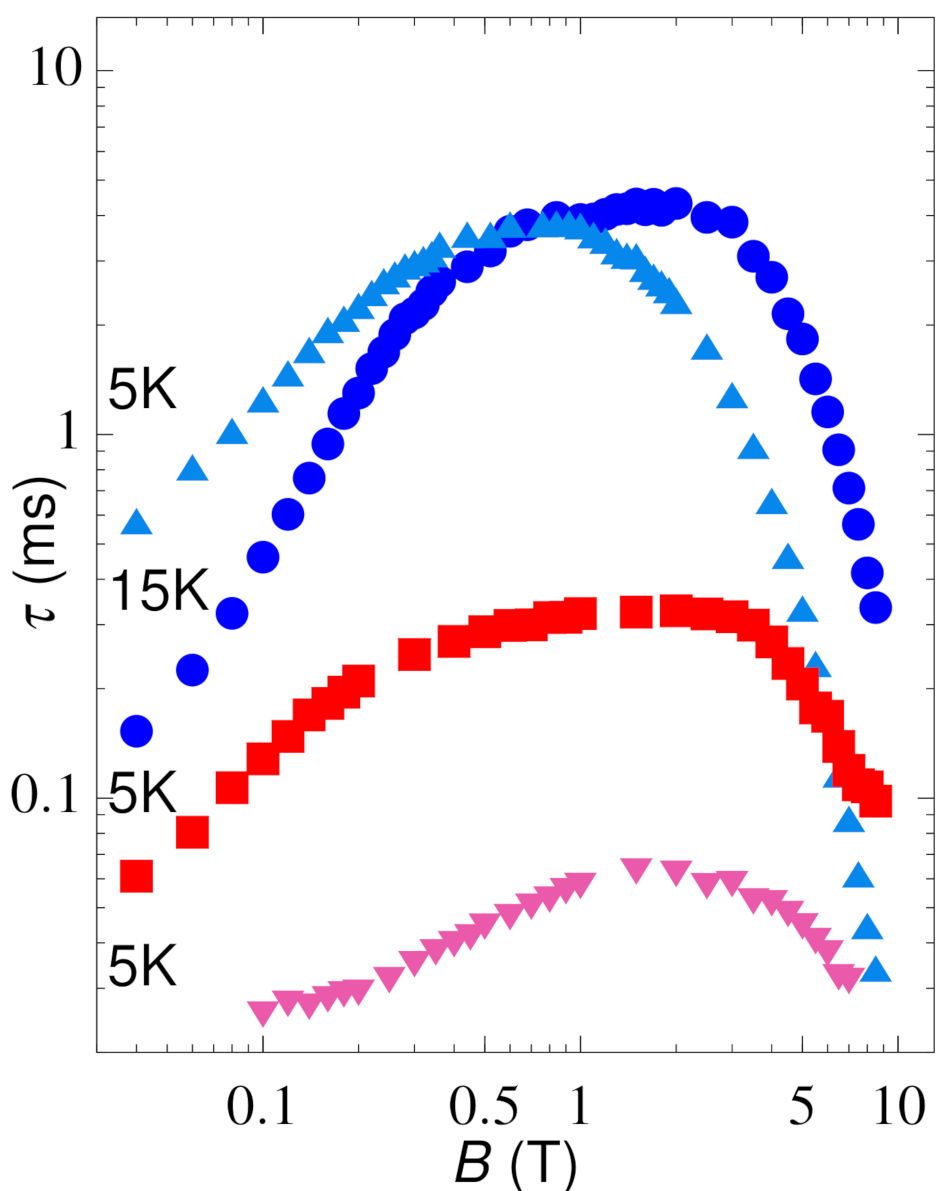}};
	\node at (6.71,3) (2) {\includegraphics[scale=1]{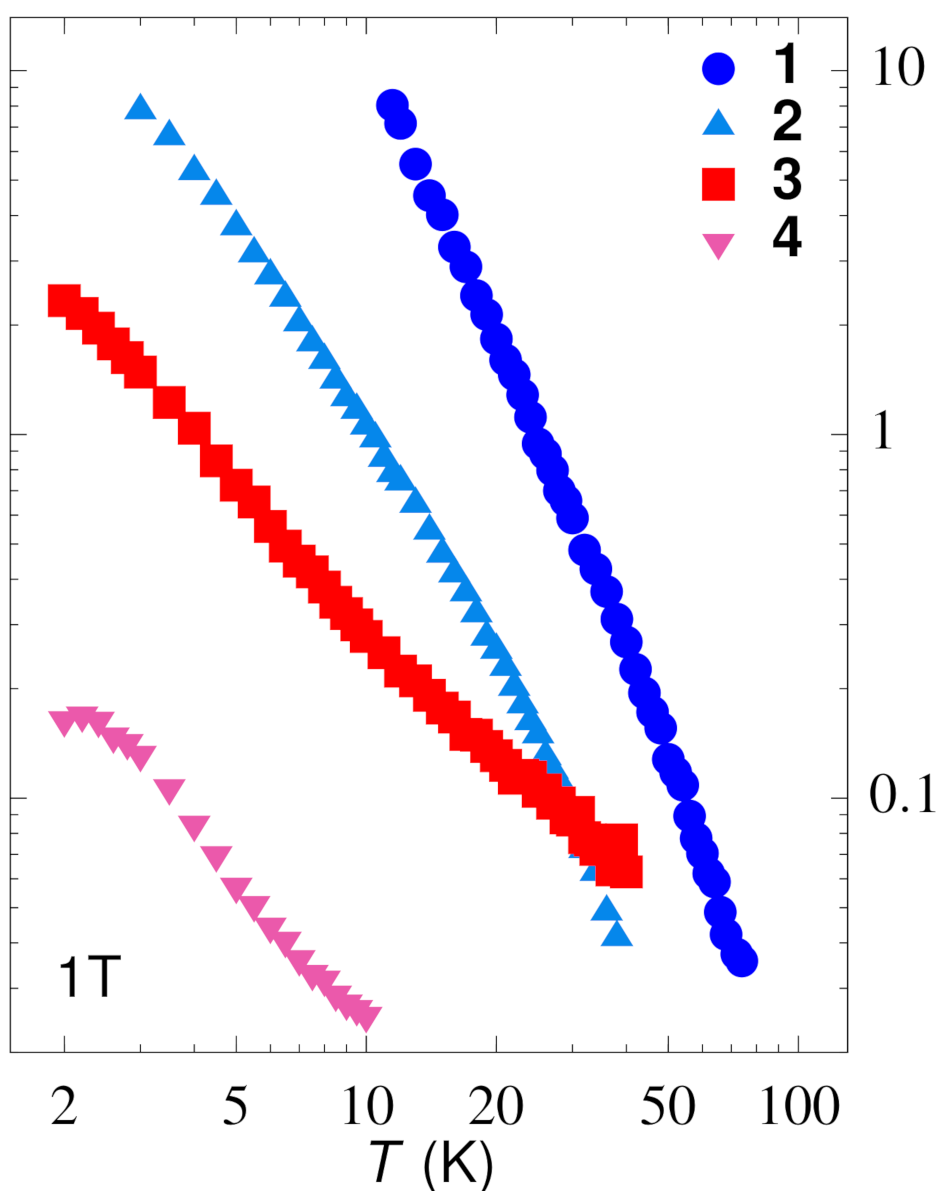}};
	
	\node at (0.3,5.45) (){\textbf{a}};
	\node at (4.5,5.45) (){\textbf{b}};
	\end{tikzpicture} 
	\caption{Spin-lattice relaxation time extracted from AC susceptometry measurements as a function of 
		the external magnetic field (left) and temperature (right) for compounds \textbf{1}-\textbf{4}. Data have been taken 
		by previous works.\cite{Atzori2016a,Atzori2017c}}
	\label{tempirelax}
\end{figure}

On the basis of the results obtained for the spin-orbit coupling projected onto the normal modes, it is now
possible to draw some considerations regarding the experimental trends of the spin-lattice relaxation. Our
methodology only includes the spin-phonon coupling originating from the modulation of the Zeeman energy by vibrations 
and, as discussed before, only the high-field regime can be considered. \autoref{tempirelax}a shows the field dependence of the 
relaxation time for the four compounds. This is characterized by a wide plateau at intermediate field values. 
We then fix the field at 1T and monitor the relaxation time as a function of temperature (\autoref{tempirelax}b). At 
this field the contribution to the relaxation coming from the hyperfine and dipolar interactions are expected to be 
reduced so that the spin-relaxation times can be correlated to our calculated spin-orbit coupling parameters.

Figures \ref{tempirelax}a and \ref{tempirelax}b show that at low temperature catecholate complexes relax slower than 
the dithiolene ones. This is consistent with the first vibrational modes structure of both compounds. Then, we note that molecules presenting hexa-coordination relax faster than those with penta geometry, 
regardless of the chemical nature of the ligands. 

Turning to the comparison of the relaxation time among iso-ligand species, we note that \textbf{2} 
relaxes faster than \textbf{1} in virtue of a much stronger spin-phonon coupling. This suggests that the spin-phonon coupling intensity and the structural 
rigidity, here assumed proportional to the frequency of the first $\Gamma$-point vibration, play together in determining the 
spin-lattice relaxation time.

This picture is confirmed by the behaviour of the relaxation times when the temperature increases and higher energy 
modes become populated. The $\tau$ of molecule \textbf{3} decays with temperature at a much slower
pace than that of \textbf{1} and \textbf{2}, and in fact there is a cross-over between the relaxation times of \textbf{2} and 
\textbf{3} at around 30 K, with another one between \textbf{1} and \textbf{3} expected at higher temperatures. This 
experimental feature correlates well with the weak spin-phonon coupling observed over a wide frequency range for \textbf{3}, 
when this is compared to that of \textbf{1} and \textbf{2}. 

Concerning compound \textbf{4}, the AC susceptibility is measurable only in a quite limited temperature range, so that 
little conclusions can be drawn. However, previous pulsed electron paramagnetic resonance (EPR) investigations of 
crystalline compounds, diluted in diamagnetic analogues, have revealed that the coherence time of \textbf{4} 
collapses at $\sim$100~K as the result of a sharp decay of $T_1$ for temperatures above 30~K.\cite{Atzori2016a} 
This observation correlates with the calculated spin-phonon coupling (\autoref{spinfonone}). In fact, spin-phonon 
couplings comparable between \textbf{3} and \textbf{4} are observed in the low-frequency range, while in the high-frequency 
one molecule \textbf{4} presents larger couplings. At high temperature, as the high-energy modes become more occupied, a significant 
enhancement of the spin-lattice relaxation in \textbf{4} is therefore expected. The amplitude of the calculated spin-phonon 
coefficients correlates with the measured temperature behaviours over a wide vibrational energy range, showing that the 
magnetization dynamics is determined by both the efficiency of the spin-phonon coupling as well as the vibrational density 
of states.

\section{Outlook}

The development of a rationale for the chemical design of new molecular qubits must proceed through the understanding of 
the correlations between the spin-phonon dynamics and the chemical identity of the molecular 
units. Our first-principles study of four different compounds made it possible to disentangle different 
contributions to the spin-phonon coupling and connect them to chemical features. The analysis of the derivatives 
of the $\mathbf{g}$ tensors highlights the importance of several factors, both intrinsic and extrinsic.
The study of the molecular averaged spin-phonon coupling $|\partial \mathbf{g}|$ identifies the 
importance of intrinsic factors such as the efficiency of the spin-orbit coupling in the electronic ground-state. This 
quantity depends dramatically on structural features, such as the coordination geometry and the ligands field 
strength. The sensibility of the spin-phonon coupling on such structural features makes the design of new ligands, 
able to stabilize specific electronic ground states, a promising approach to new molecular qubits. From a theoretical perspective 
our finding open the way to a systematic study of the connection between coordination geometry and ligands types with 
the spin-phonon coupling intensity. The correlation between structure and spin-phonon coupling is complex in 
nature and first-principles calculations are the only way to quantitatively unveil it. However, the average magnitude of the 
spin-phonon coupling has been found to correlate with experimentally accessible physical quantities as the static g-shift 
and the energy of the first excited electronic state. This suggests an easy and qualitative way to experimentally assess the 
potential of a magnetic molecule to function as a qubit.

All these intrinsic factors must be optimized together with extrinsic factors, such as 
the composition and the energy of the phonons determined by the supramolecular arrangement. Such quantities do not only 
depends on the molecular features and include the effect of the reticular environment. The suppression of intra-molecular and 
rotational contributions to the vibrations at low energy and the reduction of the phonon density of states in the same energy 
window has a remarkable effect on spin relaxation, as suggested by the behaviour of \textbf{2} at low temperature.
The spin in this molecule, by virtue of a higher rigidity, relaxes slower than in \textbf{3} and \textbf{4} at low temperature, regardless of the higher 
spin-phonon coupling. This suggests that the role of extrinsic effects is strong enough to compensate large spin-phonon couplings. 
These extrinsic factors can be tuned following two possible strategies: by stiffening the metal-ligand bonds and by tayloring
supramolecular structures where the intra-molecular contribution to the normal modes is reduced and the phononic structure is modified and shifted at higher frequencies.\cite{Yamabayashi2018} 

In conclusion, the results of our comparative first principles investigation open new pathways for the rational design of molecular qubits based 
on the combination of the coordination geometry and ligang field with crystal engineering.

\section*{Supplementary Material}
Supplementary materials contain the representation of crystallographic cells, atomically projected cartesian spin-phonon coupling coefficients, vibrational density of states, fitting of spin-phonon coupling coefficients and list of vibrational frequencies with the corresponding spin-phonon coupling norm value. 

\section*{Conflicts of interest}
There are no conflicts to declare.

\section*{Acknowledgements}
This work has been sponsored by Science Foundation Ireland (grant 14/IA/2624), italian MIUR (through Project QCNaMoS No. PRIN 2015-HYFSRT), MOLSPIN COST action CA15128 and by QuantERA European Project SUMO. Computational resources were provided by the Trinity Centre for High Performance Computing (TCHPC) and the Irish Centre for High-End Computing (ICHEC).

\end{document}